\documentclass[12pt]{article}
\usepackage{epsfig, rotating}
\def\beq{\begin{eqnarray}}
\def\eeq{\end{eqnarray}}
\usepackage{amssymb}
\usepackage{amsmath}
\begin{document}

\title{Multiple resonance compensation for 
betraton coupling and its equivalence 
with matrix method}

\author{G. De Ninno$^1$ \& D. Fanelli$^2$}
\maketitle

\begin{center}
\begin{tabular}{ll}
$^1$ & CERN PS/HP, 1211 Geneve 23, Switzerland\\
$^2$ & Dept. of Numerical Analysis and Computer Science, KTH,\\
& $\qquad$ S-100 44 Stockholm, Sweden\\
\end{tabular}
\end{center}

\begin{abstract}

Analyses of betatron coupling can be broadly divided into two categories: 
the matrix
approach that decouples the single-turn matrix to reveal the normal modes 
and the
hamiltonian approach that evaluates the coupling in terms of the action of
resonances in perturbation theory. The latter is often regarded as
being less exact but good for physical insight. The common opinion is that the
correction of the two closest sum and difference resonances to the working point
is sufficient to reduce the off-axis terms in the $4\times 4$ single-turn
matrix, but this is only partially true. The reason for this is explained,
and a method is developed that sums to infinity all coupling resonances and, in
this way, obtains results equivalent to the matrix approach. The two approaches is 
discussed with reference to the dynamic
aperture. Finally, the extension of the summation method to resonances of all
orders is outlined and the relative importance of a single resonance compared to
all resonances of a given order is analytically described as a function of the
working point.
\end{abstract}

\section{Introduction}
The example of linear betatron coupling will be used in the first instance to
demonstrate the summation of the influences of all the resonances in a given
family into a single driving term. It will then be shown that is in fact a
general result that can be applied to all linear and non-linear resonances.
\\
Analyses of betatron coupling can be broadly divided into two categories: 
the matrix approach \cite{ed+te}, \cite{talman2}, \cite{peggs}
that decouples the single-turn matrix to reveal the normal modes and the
hamiltonian approach \cite {guignard1}, \cite{phil1} 
that evaluates the coupling in terms of the action of
resonances using a perturbation method. The latter is often regarded as
being less exact but good for physical insight. The general belief is that the
correction of the two closest sum and difference resonances to the working point
should be sufficient to reduce the off-axis terms in the $4\times 4$ single-turn
matrix, but in most cases this is not successful.

\subsection{Matrix method for coupling compensation} 
The $4\times 4$ single-turn matrix in the 
presence of skew quadrupoles and/or solenoids is of the form
\begin{equation}
 {\bf T}=
\left(
\begin{array}{cc}
M&n\\
\\
m&N
\end{array}
\right),
\end{equation}
(where $M,n,m,N\in\Re^{2\times2}$).
Coupling compensation is achieved by setting the two $2\times2$ matrices 
$n$ and $m$ to zero. Due to symplecticity and periodicity of $T$ only four
free parameters (that is the strengths of four compensator units) are required.
However, this compensation is only valid at the origin of ${\bf T}$. 
\\
A transformation
can also be applied to the matrix ${\bf T}$ that decouples the linear motion, so
making it possible to describe the beam in the whole machine with the well-known
Courant and Snyder parametrisation in the transformed coordinates.

\subsection{Classical hamiltonian method for coupling compensation}
This method is based on the expansion in a Fourier series of the coupling
perturbation term in the Hamiltonian. The standard procedure is to assume that
the low-frequency components dominate the motion and that only the nearest sum
and difference resonances therefore require compensation (single resonance
compensation).
\\
The essential difference between this and the matrix approach is:
\\
\\
$~~~\bullet$~~The matrix method is exact while the hamiltonian method is
approximate;
\\
$~~~\bullet$~~A coupling compensation made by the matrix method is only valid at
one point in the ring whereas the hamiltonian method gives a global correction;
\\
$~~~\bullet$~~The matrix method leaves finite excitations in all resonances,
including those closest to the working point, whereas the hamiltonian method
leaves finite excitations only in the far resonances.
\\
\\
The reason for the two last points is that the matrix method includes all
resonances automatically and combines them in such a way that the matrix is
uncoupled at one point, while the hamiltonian method sets only the closest sum
and difference resonances to zero. If the far resonances have little effect,
then the two methods are virtually equivalent. This is however an 
uncommon situation.
\\
The logical implication is that by finding a way to sum all the resonances, the
classical hamiltonian method can be made to reproduce the results found
with the matrix method. Once this is done, the natural questions are which of
the two methods is the better for operation, and if the principle can be
extended to higher orders.
\\
\\
$\mbox{\bf {The aims of this paper are the following:}}$
\\
$~~~\bullet$~~To outline a summed resonance compensation procedure (taking
into account both the low and high-frequency part of the perturbative
Hamiltonian) for the case of linear coupling and to extend this result 
to the non-linear case (Section \ref{MRCflc});
\\
$~~~\bullet$~~To analytically compare (Section \ref{generale}) the single and 
summed resonance theories 
pointing out some general results that can be obtained using 
the analytical expression of the generalized driving term; 
\\
$~~~\bullet$~~To numerically compare the single and the summed resonance 
compensations for the linear coupling (the latter
is shown to be
equivalent to the matrix compensation) using a 4D coupled Henon map
\cite{Henon} (Section \ref{chm});

\section{Multiple resonance compensation for linear coupling}
\label{MRCflc}
The starting point for this analysis is taken from \cite{guignard1},
with the initial assumptions that:
\\
$~~~\bullet$~~the perturbative Hamiltonian is calculated at $\theta=0$. Since the
origin is an arbitrary choice, this is not a restriction;
\\
$~~~\bullet$~~solenoid fields are absent. The presence of solenoid fields 
does not change the argument but by omitting them 
the resulting equations become more transparent.
They will be added at the end.
\\
\\
The linear coupling compensation using the notation of \cite{guignard1}
requires (without any approximation for 
removing the high frequency part):
\begin{equation}
\label{diffesum}
\left\{
\begin{array}{l}                                        
\displaystyle{\sum_{p=-\infty}^{+\infty}h^{(2)}_{1010-p}=0}~~~~~~~~
\mbox{sum resonance}
\\
\\
\displaystyle{\sum_{p=-\infty}^{+\infty}h^{(2)}_{1001-p}=0}~~~~~~~~
\mbox{difference resonance}.
\end{array}
\right.
\end{equation}

\subsection{Detailed derivation for difference resonances}
Consider first the treatment of the difference resonances and express 
(\ref{diffesum}) explicitly as
\\
\begin{equation}
\sum_{p=-\infty}^{+\infty}h^{(2)}_{1001-p}\equiv C^-_{\infty}=
\sum_{p=-\infty}^{\infty}\int_0^{2\pi}A(\theta)e^
{\displaystyle{-i(Q_x-Q_y-p)\theta}}
\mbox{d}~\theta
\end{equation}
where
\begin{equation}
A(\theta)=\frac{1}{4\pi R}\sqrt{\beta_x(\theta)\beta_y(\theta)}
e^{\displaystyle{i[\mu_x(\theta)-\mu_y(\theta)]}}k(\theta),
\end{equation}
\begin{equation}
k(\theta)=\frac{R^2}{B\rho}\frac{\partial
B_x}{\partial x},
\end{equation}
$\theta=s/R$ is the coordinate along the ring,
$Q_{x,y}$ are the horizontal and vertical tunes, $\mu_{x,y}$ are the
horizontal and vertical phase advances, $\beta_{x,y}$ are the horizontal and
vertical beta functions and $R$ is the radius of the ring.
\\
Suppose $k(\theta)$ is different from zero and constant in $j$ short intervals 
(i.e. the
regions occupied by the sources of coupling and by possible correctors) $[\theta_i,
\theta_i+\Delta\theta_i]$ in which $A(\theta)$ can be considered approximately
constant (thin lens approximation) \footnote{In a real machine $A(\theta)$
will vary slowly or, at least, it will be possible to cut the elements into
short enough pieces that $A(\theta)$ can be considered as constant over all
sub-elements to any desired degree of accuracy.}:
\begin{equation}
C^-_{\infty}=\sum_{i=1}^{j}(\Delta C^-_{\infty})_i.
\end{equation}
\\
Now define $\Delta^-\equiv Q_x-Q_y$ and compute $(\Delta C^-_{\infty})_i$,
the contribution to $C^-_{\infty}$ from the $i-th$ sub-element,
\begin{equation}
(\Delta
C^-_{\infty})_i=\sum_{p=-\infty}^{+\infty}\int_{\theta_i}^
{\theta_i+\Delta\theta_i}A(\theta)e^{\displaystyle{-i[\Delta^--p]\theta}}
\mbox{d}\theta,
\end{equation}
which, when integrated gives,
\begin{equation}
(\Delta
C^-_{\infty})_i\simeq
A(\theta_i)
\sum_{p=-\infty}^{+\infty}\frac{i}{\Delta^--p}\left[e^{\displaystyle{-i[\Delta^-
-p](\theta_i+\Delta\theta_i)}}-e^{\displaystyle{-i[\Delta^--p]\theta_i}}\right].
\end{equation}
The summation can be redefined making use of the shift $[\Delta^-]$ (the closest
integer to $\Delta^-$) so that $p=[\Delta^-]+k$
where $k$ is an integer. In the limit $(\Delta^- - [\Delta^-])\Delta\theta_i
\ll 1$, the previous expression becomes\footnote{In the expression 
(\ref{otto})
the term $\displaystyle{e^{-i(\Delta^- - [\Delta^-])\Delta\theta}}$ is expanded 
to the zero order while the terms $\displaystyle{e^{ik\Delta\theta}}$ 
with $k\simeq (\Delta^- - [\Delta^-])$ are not expanded. 
This assumption, supported "a posteriori" by the accuracy of the final result,
is based on the fact that the contribution of the higher order terms to the
sum of the series is negligible.} 
\begin{eqnarray}
\label{otto}
(\Delta
C^-_{\infty})_i&\simeq&A(\theta_i)
e^{\displaystyle{-i(\Delta^- - [\Delta^-])\theta_i}}\sum_{k=-\infty}^{+\infty}
\frac{i}{(\Delta^- - [\Delta^-])-k}
\left[e^{\displaystyle{ik(\theta_i+\Delta\theta_i)}}-
e^{\displaystyle{ik\theta_i}}\right]=\nonumber\\
&=&
A(\theta_i)
e^{\displaystyle{-i(\Delta^- - [\Delta^-])\theta_i}}
\sum_{k=-\infty}^{+\infty}\frac{1}{(\Delta^- - [\Delta^-]) -k}\{
\sin(k\theta_i)-\sin[k(\theta_i+\Delta\theta_i)]+\nonumber\\
&+&i\left[\cos[k(\theta_i+\Delta\theta_i)]-\cos(k\theta_i)\right]\}.
\end{eqnarray}
To sum these series first we rewrite them in a more suitable form 
($x=\theta_i$ or $x=\theta_i+\Delta\theta_i$) valid when $0< x < 2 \pi$
\cite{grads}:
\begin{eqnarray}
\label{serie1}
\sum_{k=-\infty}^{+\infty}\frac{\sin(kx)}{(\Delta^- - [\Delta^-])-k}=
-\frac{\pi\sin[(\Delta^- - [\Delta^-])(\pi-x)]}{\sin((\Delta^- 
- [\Delta^-])\pi)};
\end{eqnarray}

\begin{eqnarray}
\label{serie2}
\sum_{k=-\infty}^{+\infty}\frac{\cos(kx)}{(\Delta^- - [\Delta^-])-k}=
\frac{\pi\cos[(\Delta^- - [\Delta^-])(\pi-x)]}
{\sin((\Delta^- - [\Delta^-])\pi)}.
\end{eqnarray}
The application of (\ref{serie1}) and (\ref{serie2}) to (\ref{otto}) then gives
\begin{eqnarray}
\label{sommainfinita}
(\Delta C^-_{\infty})_i&=&A(\theta_i)
e^{\displaystyle{-i(\Delta^- - [\Delta^-])\theta_i}}
\frac{\pi}{\sin((\Delta^- - [\Delta^-])\pi)}
\{(\sin[(\Delta^- - [\Delta^-])(\pi-\theta_i-\Delta\theta_i)]+\nonumber\\
&&\nonumber\\
&-&\sin[(\Delta^- - [\Delta^-])(\pi-\theta_i)]+
i[\cos[(\Delta^- - [\Delta^-])(\pi-\theta_i-\Delta\theta_i)]+\nonumber\\
&&\nonumber\\
&-&\cos[(\Delta^- - [\Delta^-])(\pi-\theta_i)]]\}=\nonumber\\
&&\nonumber\\
&=&A(\theta_i)
e^{\displaystyle{-i(\Delta^- - [\Delta^-])\theta_i}}
\frac{2\pi}{\sin((\Delta^- - [\Delta^-])\pi)}
\sin[(\Delta^- - [\Delta^-])\frac{\Delta\theta_i}{2}]\cdot\nonumber\\
&&\nonumber\\
&\cdot&e^{\displaystyle{-i[(\Delta^- - [\Delta^-])
(\pi-\theta_i-\frac{\Delta\theta_i}{2})]}}.
\end{eqnarray}
After expressing $A(\theta_i)$ explicitly, $(\Delta C^-_{\infty})_i$ becomes 
\begin{equation}
\label{multicom}
(\Delta C^-_{\infty})_i
\simeq-
\frac{k_i}{4R}\frac{(\Delta^- - [\Delta^-])\Delta\theta_i}
{\sin((\Delta^- - [\Delta^-])\pi)}
\sqrt{\displaystyle{\beta_x(\theta_i)\beta_y(\theta_i)}}~
e^{\displaystyle{i\left(\mu_x(\theta_i)-\mu_y(\theta_i)-
(\Delta^- - [\Delta^-])\pi
\right)}}
\end{equation}
Equation (\ref{multicom}) can be summed directly for all the elementary elements
in the ring to give the coupling coefficient $C^-_{\infty}$ for the combined
influence of all the linear difference resonances.

\subsection{Extension to sum resonances and solenoids}
\label{extension}
For the sum resonances the procedure is unchanged and the
formal result is the same, but with the following substitutions:
\begin{equation}
\mu_x(\theta_i)-\mu_y(\theta_i)\longrightarrow \mu_x(\theta_i)+\mu_y(\theta_i)
\end{equation}
\begin{equation}
\Delta^-\longrightarrow \Delta^+\equiv Q_x+Q_y.
\end{equation} 
\\
In presence of a uniform solenoidal field the summed resonance driving term can again 
be written in the form
\begin{equation}
C_{\infty}^{\pm}=\sum_{p=-\infty}^{+\infty}\int_0^{2\pi}A^{\pm}(\theta)
e^{\displaystyle{-i(\Delta^{\pm}-p)\theta}}\mbox{d}\theta
\end{equation}
where now 
\begin{equation}
A^{\pm}(\theta)=\frac{1}{4\pi R}\sqrt{\beta_x(\theta_i)\beta_y(\theta_i)}
RS\left[\left(\frac{\alpha_x}{\beta_x}-\frac{\alpha_y}{\beta_y}\right)
-
i\left(\frac{1}{\beta_x}\mp\frac{1}{\beta_y}\right)\right]
e^{\displaystyle{i[\mu_x(\theta)\pm\mu_y(\theta)]}}
\end{equation}
($k=0$, $S=\displaystyle{\frac{R}{2B\rho}B_{\theta}}$).
The same procedure as before yelds
the following expressions for $(\Delta C^+_{\infty})_i$ and 
$(\Delta C^-_{\infty})_i$:
\begin{eqnarray}
(\Delta C^{\pm}_{\infty})_i&=&
-\frac{S_i}{2}\frac{\sin((\Delta^{\pm} - [\Delta^{\pm}])
\Delta\theta_i/2)}{\sin((\Delta^{\pm} - [\Delta^{\pm}]) \pi)}
\left[\left(\frac{\alpha_x}{\beta_x}-\frac{\alpha_y}{\beta_y}\right)-
i\left(\frac{1}{\beta_x}\mp\frac{1}{\beta_y}\right)\right]_{\theta=\theta_i}
\cdot\nonumber\\
&&\nonumber\\
&\cdot&\sqrt{\beta_x(\theta_i)\beta_y(\theta_i)}
e^{\displaystyle{i[\mu_x(\theta_i)\pm\mu_y(\theta_i)
-(\Delta^{\pm} - [\Delta^{\pm}])\pi]}}.
\end{eqnarray}
\\

\subsection{Extention to the nonlinear case}
It is clear from Section \ref{extension} that the procedure for summing the
resonances is, in fact, independent of the detailed form of the term 
$A(\theta)$ and that, with the general form of $A(\theta)$ from
\cite{guignard2}, the method can be extended to the nonlinear case.
\\
The driving term of a given resonance $(n_1,n_2)$ of order $N=n_1+n_2$
for the single and summed resonance theories are (respectively)
\\
\\
\\

\begin{equation}
 C_{n_1,n_2,p}=
\frac{1}{\pi(2R)^{N/2}|n_1|!|n_2|!}\int_0^{2\pi}\beta_x^{|n_1|/2}
\beta_y^{|n_2|/2}
e^{i\left[n_1\mu_x+n_2\mu_y-(\Delta-p)\theta\right]}k d\theta
\end{equation}

and

\begin{equation}
C_{n_1,n_2,\infty}=
\frac{\pi(\Delta-[\Delta])}{\sin\left[\pi(\Delta-[\Delta])\right]}
\frac{1}{\pi(2R)^{N/2}|n_1|!|n_2|!}\int_0^{2\pi}\beta_x^{|n_1|/2}
\beta_y^{|n_2|/2}
e^{i\left[n_1\mu_x+n_2\mu_y-(\Delta-[\Delta])\pi\right]}k d\theta
\end{equation}
where\footnote{In the following formulas the partial derivatives are evaluated in $x=y=0$}:

\begin{eqnarray}
k&=&(-1)^{\frac{|n_2|+2}{2}}\frac{R^2}{2|B\rho|}
[(-1)^{\frac{|n_2|+2}{2}}(\frac{\partial^{(N-1)}B_y}
{\partial x^{(N-|n_2|-2)}\partial y^{(|n_2|+1)}}
-\frac{\partial^{(N-1)}B_x}{\partial x^{(N-|n_2|-1)}\partial y^{|n_2|}})
+\nonumber\\
&+&
\frac{\partial^{(N-2)}B_s}{\partial x^{(N-|n_2|-2)}\partial y^{|n_2|}}(|n_1|
\frac{\alpha_x}{\beta_x}-|n_2|\frac{\alpha_y}{\beta_y})
-i\frac{\partial^{(N-2)}B_s}{\partial x^{(N-|n_2|-2)}\partial y^{|n_2|}}
(\frac{n_1}{\beta_x}-\frac{n_2}{\beta_y})]
\end{eqnarray}

for $|n_2|$ even, $N \ge 3$ and $1 \le |n_2| \le (N-2)$;

\begin{eqnarray}
k&=&(-1)^{\frac{|n_2|-1}{2}}\frac{R^2}{2|B\rho|}
[(-1)^{\frac{|n_2|-1}{2}}(\frac{\partial^{(N-1)}B_y}
{\partial x^{(N-|n_2|-2)}\partial y^{(|n_2|+1)}}
-\frac{\partial^{(N-1)}B_x}{\partial x^{(N-|n_2|-1)}\partial y^{|n_2|}})
+\nonumber\\
&+&
\frac{\partial^{(N-2)}B_s}{\partial x^{(N-|n_2|-2)}\partial y^{|n_2|}}(|n_1|
\frac{\alpha_x}{\beta_x}-|n_2|\frac{\alpha_y}{\beta_y})
-i\frac{\partial^{(N-2)}B_s}{\partial x^{(N-|n_2|-2)}\partial y^{|n_2|}}
(\frac{n_1}{\beta_x}-\frac{n_2}{\beta_y})]
\end{eqnarray}

for $|n_2|$ odd, $N \ge 3$ and $1 \le |n_2| \le (N-2)$.

It is customary to use the symbols $C$ for the coupling driving terms and $K$
for the higher-order non linear driving terms. There is also often a factor of 2
between the two definitions ($K=C/2$ for a given resonance).
Since this report concentrates on coupling the "$C$"-styled definition has been
extended to all the cases. 
\\
\section{General results}
\label{generale}
This section is dedicated to pointing out some of the general consequences 
of the summed resonance theory.

\subsection{Analytic comparison of the influence of single and summed 
resonances}
\label{anco}
It is interesting to compare the contribution to the coupling 
excitation from all resonances to that of the closest 
single resonance.
The single resonance driving term \cite{guignard1} reads 
\begin{equation}
C_{p}=\int_0^{2\pi}A(\theta)
e^{\displaystyle{-i(\Delta -p)\theta}}\mbox{d}\theta
\end{equation} 
where now $p$ is the closest integer to $\Delta$.
\\
\begin{figure}[ht!]
\centering
\epsfig{file=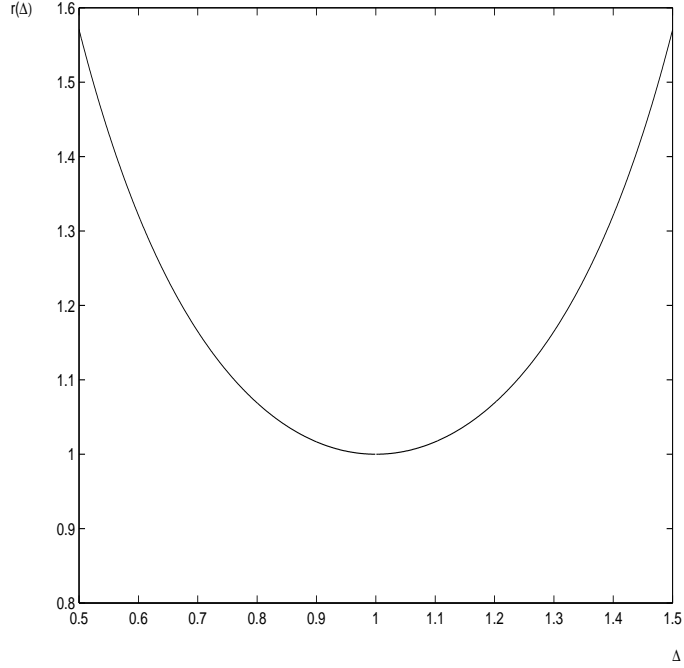, height=9truecm,width=9truecm}
\caption{\label{modulii.eps}  Ratio r between $\|\Delta C^{\pm}_{\infty}\|$ 
and $\|\Delta C^{\pm}_{1}\|$ versus $\Delta^{\pm}$.}
\end{figure}

Using the thin lens approximation the $i-th$ contribution to $C_{p}$ can
be written
\begin{eqnarray}
\label{guigny}
(\Delta C_{p})_i&\simeq&A(\theta_i)\int_{\theta_i}^{\theta_i+\Delta\theta_i}
e^{\displaystyle{-i(\Delta-p)\theta}}\mbox{d}\theta=\nonumber\\
&&\nonumber\\
&=&A(\theta_i)\frac{i}{|\Delta^{\pm}-p|}e^{\displaystyle{-i(\Delta
-p)\theta_i}}\left[\cos[(\Delta -p)\Delta\theta_i]-1-
i\sin[(\Delta -p)\Delta\theta_i]\right]\simeq\nonumber\\
&&\nonumber\\
&\simeq&\frac{k_i}{\pi(2R)^{N/2}|n_1|!|n_2|!}\Delta\theta_i 
(\beta_x)_i^{|n_1|/2}
(\beta_y)_i^{|n_2|/2}
e^{i[n_1(\mu_x)_i+n_2(\mu_y)_i-(\Delta -p)\theta_i]}.
\end{eqnarray}
Fig. (\ref{modulii.eps}) shows the ratio between 
$||(\Delta C_{\infty}^{\pm})_i||$ and 
$||(\Delta C_{p}^{\pm})_i||$ varying the distance from the resonance
($[\Delta^{\pm}]=1$).
\\
The formulae (\ref{multicom}) and (\ref{guigny}) give the same result 
for the modulus of the driving terms when
exactly on
resonance. A difference between (\ref{multicom}) and (\ref{guigny}) 
increases (approximatively) quadratically as one moves
away from $\Delta$ integer. Agreement between the two formalism can be therefore
expected only if the working point is close enough to the resonance to be
compensated \footnote{Note that the phase terms are different even when exactly
on resonance.}.
\\
However, this is not usually the case if the aim is the full 
compensation of
the linear coupling, that is both the sum and difference resonances. In most 
pratical cases, the working point is close to the difference resonance and
relatively distant from the sum resonance.
\\

\subsection{Closed-orbit distortion from a dipole kick}
The equation (\ref{multicom}) can be applied to the resonance
family
\begin{equation}                                    
\displaystyle{Q_z=p}~~~~~~~~
\end{equation}
where $z\equiv x,y$. This leads to the expressions for the
closed-orbit distortion due to a dipole kick and links the single resonance 
theory \cite{guignard1}, \cite{guignard2} to the integrated theory of 
Courant and Snyder \cite{cou+sny}.
\\ 
In this case 
\begin{equation}
C_{\infty}=
\sum_{p=-\infty}^{+\infty}
\int_{0}^{2\pi} A(\theta)e^{\displaystyle{-i(Q_z-p)\theta}}\mbox{d}\theta
\end{equation}
where now \cite{guignard2} 
\begin{equation}
A(\theta)=\frac{1}{2^{3/2}\pi R^{1/2}}\sqrt{\beta_z(\theta)}e^
{\displaystyle{i\mu_z(\theta)}}\frac{\Delta B}{B\rho}
\end{equation}  
($\Delta B/B\rho$ is the dipole error).
\\
For a localized error of length $\Delta l$ (thin lens approximation):
\begin{eqnarray}
\label{dipkick}
(\Delta C_{\infty})_i&=&-\frac{1}{2^{3/2} R^{1/2}}\frac{Q_z-[Q_z]}
{\sin(\pi(Q_z-[Q_z]))}\sqrt{\beta_z}
\frac{\Delta l \Delta B}{B \rho}
e^{\displaystyle{i(\mu_z(\theta)-(Q_z-[Q_z])\pi)}}=\nonumber\\
&&\nonumber\\
&=&
-\frac{1}{2^{3/2} R^{1/2}} \frac{Q_z-[Q_z]}{\sin(\pi Q_z)}
\frac{\Delta l \Delta B}{B \rho} \sqrt{\beta_z} 
e^{\displaystyle{i(\mu_z(\theta)-Q_z\pi)}}.
\end{eqnarray}
Comparing equation 
(\ref{dipkick}) to the closed-orbit distortion (at the origin) due
to a kick occurring at a given position $\theta$
\begin{equation}
(z)_{\theta=0}=\frac{\sqrt{\beta_z(0)}}{2\sin(\pi Q_z)}\sqrt{\beta_z(\theta)}
\frac{\Delta B \Delta l}
{B\rho}\cos(\mu_z(\theta) -\pi Q_z)
\end{equation}
shows that the normalized orbit distortion differs from 
$\mbox{Re}(\Delta C_{\infty})_i$ by only a constant,
\begin{equation}
\left(\frac{z}{\sqrt{\beta_z}}\right)_{\theta=0}\equiv 
\frac{1}{2^{-1/2} R^{-1/2}(Q_z-[Q_z])}
~\mbox{Re}(\Delta C_{\infty})_i
\end{equation}

\subsection{Betatron amplitude modulation}
Applying the same procedure to the resonance family
\begin{equation}
Q_z=2p.
\end{equation}
one gets the modulation of the betatron function due to a small gradient error
occurring at a given position $\theta$.
\\
In this case
\begin{equation}
C_{\infty}\equiv=
\sum_{p=-\infty}^{+\infty}\int_{0}^{2\pi} A(\theta)e^{\displaystyle
{i(2Q_z-p)\theta}}\mbox{d}\theta
\end{equation}
with
\begin{equation}
A(\theta)=\frac{1}{4\pi R}\beta_z(\theta) e^{\displaystyle{2i\mu_z(\theta)}}
\frac{R^2}{B\rho}\frac{\partial B_y}{\partial x}.
\end{equation}
This gives
\begin{eqnarray}
(\Delta C_{\infty})_i&=&-\frac{1}{2}\frac{Q_z-[Q_z]}
{\sin(2\pi(Q_z-[Q_z]))}\beta_z(\theta)
\frac{\Delta l}{B\rho}\frac{\partial B_y}{\partial x}
e^{\displaystyle{i(2\mu_z(\theta)-2(Q_z-[Q_z])\pi)}}=\nonumber\\
&&\nonumber\\
&=&
-\frac{1}{2} \frac{Q_z-[Q_z]}{\sin(2\pi Q_z)} 
\beta_z(\theta) 
\frac{\Delta l}{B\rho}\frac{\partial B_y}{\partial x}
e^{\displaystyle{i(2\mu_z(\theta)-2Q_z\pi)}}.
\end{eqnarray}
The last expression coincides with
the modulation of the beta function (at the origin)
\begin{equation}
\left(\Delta \beta_z\right)_{\theta=0}=
\frac{\beta_z(0)}{\sin(2\pi Q_z)}\beta_z(\theta)
\frac{\Delta l}{B\rho}\frac{\partial B_y}{\partial x}
\cos(2(\mu_z(\theta)-Q_z\pi))
\end{equation}
and shows that the normalized modulation differs from 
$\mbox{Re}(\Delta C_{\infty})_i$ by only a constant, 
\begin{equation}
\left(\frac{\Delta \beta_z}{\beta_z}\right)_{\theta=0}\equiv
\frac{1}{(Q_z-[Q_z])/2}
~\mbox{Re}(\Delta C_{\infty})_i
\end{equation}

\subsection{Comments}
The summed resonance driving terms lead naturally to a new definition of
bandwidth. However, pratical differences maybe not so clearly observed. 
When the working point is close to a resonance, the bandwidth is important but
the single and summed theories do not differ much. When the working point is far from the
resonance, the bandwidth is unimportant and any widening may go unnoticed.
\\
There is also the more academic point that the magnitude of the summed resonance
driving term is dependent on the azimuthal position in the machine. The closed
orbit is a very good example of this. If there is a closed bump in the orbit,
then the $\mbox{Re}(C_{1,0,\infty})$ is zero outside the bump and finite inside.
The implication is that the beam is responding to standing waves from each
member of the resonance family. The summed response is the sum of these
standing waves. When inside a bandwidth the growth rates are related to the
standing wave amplitude and varying according to the position around the
machine.

\section{The coupled Henon map}
\label{chm}
In this section, the summed resonance approach is shown to be equivalent to the 
matrix approach and both are compared to the single-resonance compensation
by performing a numerical analysis on the so-called Henon map
\cite{Henon}: a hyper-simplified lattice model \footnote{A linear
lattice model containing only one sextupolar kick.} whose phase-space
trajectories show some of the expected characteristic of a realistic lattice 
map (nonlinearities, regions of regular and stochastic motion etc.) 
 In this application the linear coupling is generated 
and corrected by 1+4 
thin skew quadrupoles \footnote{Lattices with only solenoids or with 
both types of coupling elements give the same kind of results.}. 
\\
The global compensation of the coupling resonances (at $\theta$=0) is achieved
if
\begin{equation}
\label{globmenopiu}
\left\{
\begin{array}{l}
\displaystyle{\sum_{j=1}^5 [\mbox{Re}(\Delta C^-_{\infty})
+\mbox{i}\mbox{Im}(\Delta C^-_{\infty})]=0}
\\
\\
\displaystyle{\sum_{j=1}^5 [\mbox{Re}(\Delta C^+_{\infty})
+\mbox{i}\mbox{Im}(\Delta C^+_{\infty})]=0}.
\end{array}
\right.
\end{equation}
\\
The compensation for both the sum and the difference resonance is obtained
solving the system of 4-equations (for the 4 unknowns $k_i$) given by  
(\ref{globmenopiu}).
\\ 
\begin{table}[h]
\begin{center}
\vspace{5mm}
\begin{tabular}{|c| |c| |c| |c| }
\hline 
$k~(\mbox{m}^{-2})$&Matrix&Summed&Single\\
\hline\hline
$k_1$ (source)&0.5&0.5&0.5\\
\hline
$k_2$&-0.051&-0.050&0.559\\
\hline
$k_3$&0.034&0.033&0.554\\
\hline
$k_4$&-0.319&-0.313&0.476\\
\hline
$k_5$&-0.275&-0.275&0.117\\
\hline
\end{tabular}
\end{center}
\caption{Compensator strengths ($k_{2-5}$) in presence of the coupling source
$k_1$ using the single-turn matrix compensation and the summed and single
resonance compensations.}
\end{table}
\\
Table 1. shows a comparison between the strengths of the 4 correctors ($k_{2-5}$)
when compensating the single-turn matrix, the two infinite families of sum and 
difference
resonances (for the same $\theta=0$) and the closest sum and difference
resonances to the working point. The single-turn matrix compensation has been
performed by means of the MAD program \cite{mad} while the (single and 
multiple) resonance compensation has been obtained making use of the AGILE
program \cite{phil2} in which the formula (\ref{multicom}) has been implemented.
\\
The single-turn matrix (in $\theta=0$) in presence of the coupling source $k_1=0.5~
\mbox{m}^{-2}$ (no compensation) has non zero off-axis $2\times2$ sub-matrices given
by
\begin{equation}
{\bf T}=
\left(
\begin{array}{cc}
M&n\\
\\
m&N
\end{array}
\right)=
\left(
\begin{array}{cccc}
&&0.49&0.49\\
&&0.02&0.02\\
-0.23&-0.24&&\\
-0.01&-0.01&&
\end{array}
\right)
\end{equation}   
while the residual values of $n$ and $m$ after the single resonance
compensation ($C^+=C^-=0$) are given by
\begin{equation}
{\bf T}=
\left(
\begin{array}{cccc}
&&-0.11&-4.67\\
&&0.02&0.15\\
0.03&-4.57&&\\
0.03&-0.09&&
\end{array}
\right).
\end{equation} 
The off-axis terms are in fact larger after the compensation then before.
This is explained by the influence of the far resonances that can not be 
neglected, to get a satisfactory coupling compensation in this case.
\\
The same conclusion may be drawn by looking at the driving terms 
of the closest 
sum and difference resonances before the correction
\begin{equation}
|C^+|=0.0628  ~~~~~~|C^-|=0.0628
\end{equation}
and afterwards
\begin{equation}
|C^+|=0.0207  ~~~~~~|C^-|=0.1018.
\end{equation}
Last two equations show that the sets of the uncompensated sum and difference
resonances have a "weight" comparable (larger in the case of the difference
resonance) to the ones  compensated. 
\\
\\
The quantitative difference between the two approaches can be better
investigated by means of a tracking analysis.
\\
In the following, the results
from stability and footprint diagrams as well as the calculation of the
dynamic aperture for the compensated Henon map are shown.

\subsection{Stability and footprint diagrams}
A stability diagram and the related frequency diagram can be obtained by
the following procedure: for each initial condition inside a given grid in
the physical plane $(x,y)$ ($p_x=p_y=0$), the symplectic map representing the
lattice is iterated over a certain number of turns. If the orbit is still 
stable after the
last turn, the nonlinear tunes can be calculated using one of the methods
described in \cite{tune}.
In the stability diagram the stable initial conditions are plotted whereas in
the frequency diagram are represented the corresponding tunes. The insertion in the
frequency (footprint) diagram of the straight lines representing the resonant 
conditions up to a certain order makes it possible to visualize the
excited resonances close to the reference orbit.
\\

Figs. (5) - (10) show the stability and 
frequency diagrams for the uncoupled
Henon map, after the single resonance compensation and after the summed
resonance one.
\\
The comparison points out that the summed resonance compensation allows a 
more efficient
restoration of the uncoupled optics. It is significant the
analysis of the degree of excitation relative to the 
resonances (3,-6), (1,-4) and (2,-5) for the two different compensation
approaches.
\\

Using the perturbative tools of normal forms \cite{normform}
one can calculate the value of the first resonant coefficient 
(leading term) in the interpolating Hamiltonian for the considered resonances.
The leading term can be considered as a "measure" of the resonance excitation. 
It can be shown
\cite{deto} that in absence of coupling the leading term of the resonances 
(3,-6) and (1,-4) is different from zero (first order excitation) whereas the 
one of the resonance (2,-5) is zero (second order excitation). 
The strength of the coupling (that is, in the considered case, the strength of 
the residual coupling after the compensations) is proportional to the growth of
the leading term of the first order non-excited resonances and to the decrease 
of the leading term of the other ones. 
\\
The resonance degree of the excitation varying the compensation approach can
be better visualized ploting the network of the resonances and their widths 
inside the stability domain. The analysis of Figs.
(2)-(4) confirms that the SR method is
characterized by 
a residual coupling considerably stronger than the one left 
by the MR compensation.  
\\
The same conclusion can be drawn the following topological
argument. A trace of the presence of linear coupling in a nonlinear lattice is
the spliting of the resonant channels in correspondence of the crossing
points (multiple resonance condition in the tune space). This phenomenon is
evident only in the case of the MR compensation  (see the central part of Fig.
(4)).

\begin{figure}[ht!]
\centering
\epsfig{file=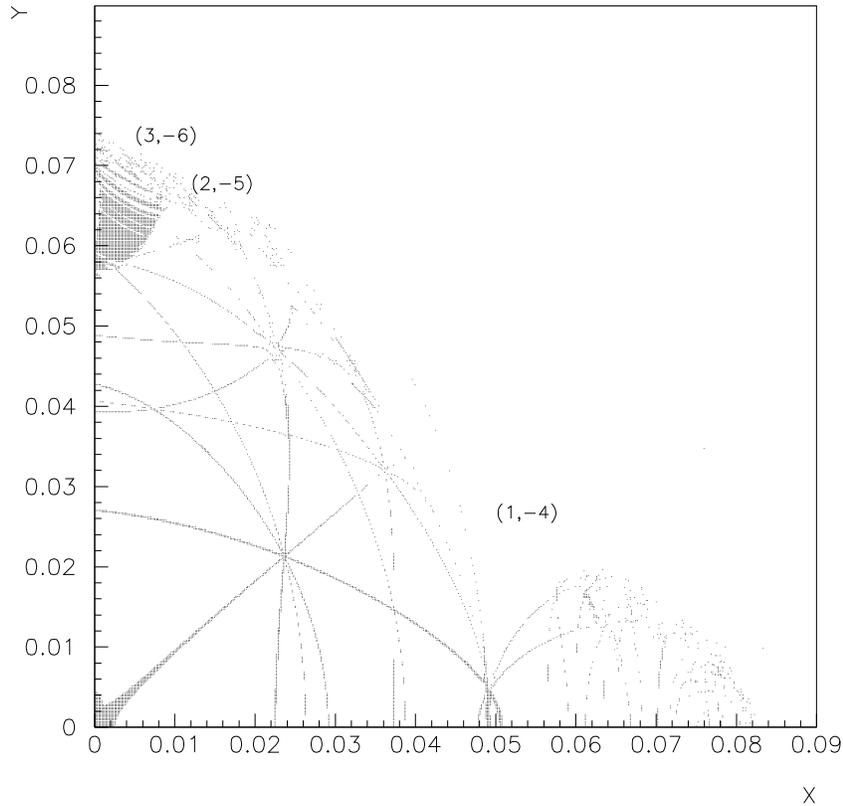, height=12truecm,width=12truecm}
\caption{Network of resonances of the uncoupled Henon
map.}
\end{figure}

\begin{figure}[ht!]
\centering
\epsfig{file=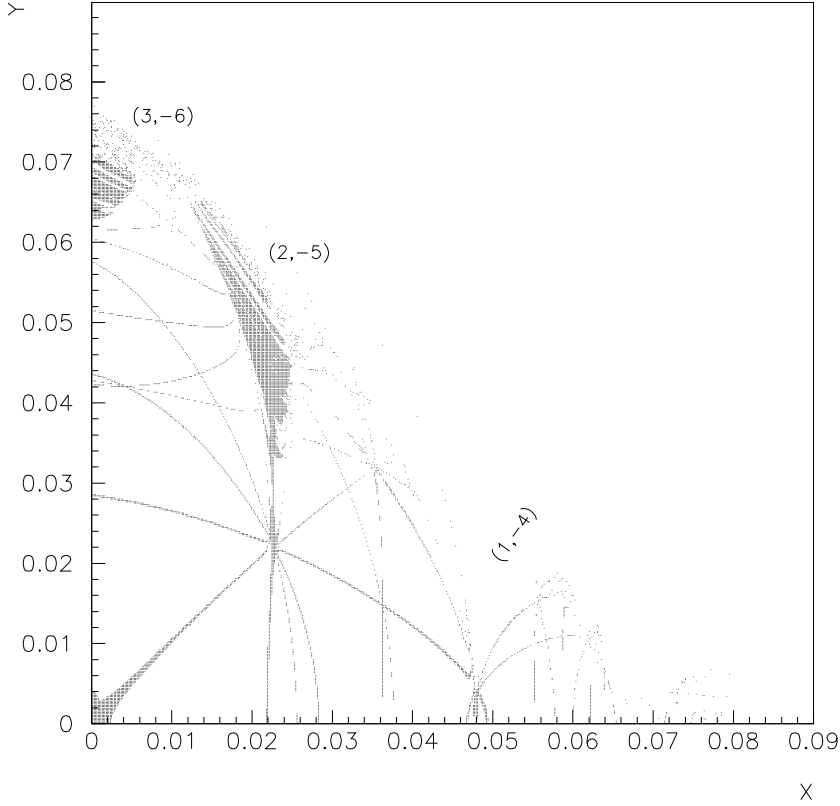, height=12truecm,width=12truecm}
\caption{Network of resonances after the summed resonance
compensation.}
\end{figure}  

\begin{figure}[ht!]
\centering
\epsfig{file=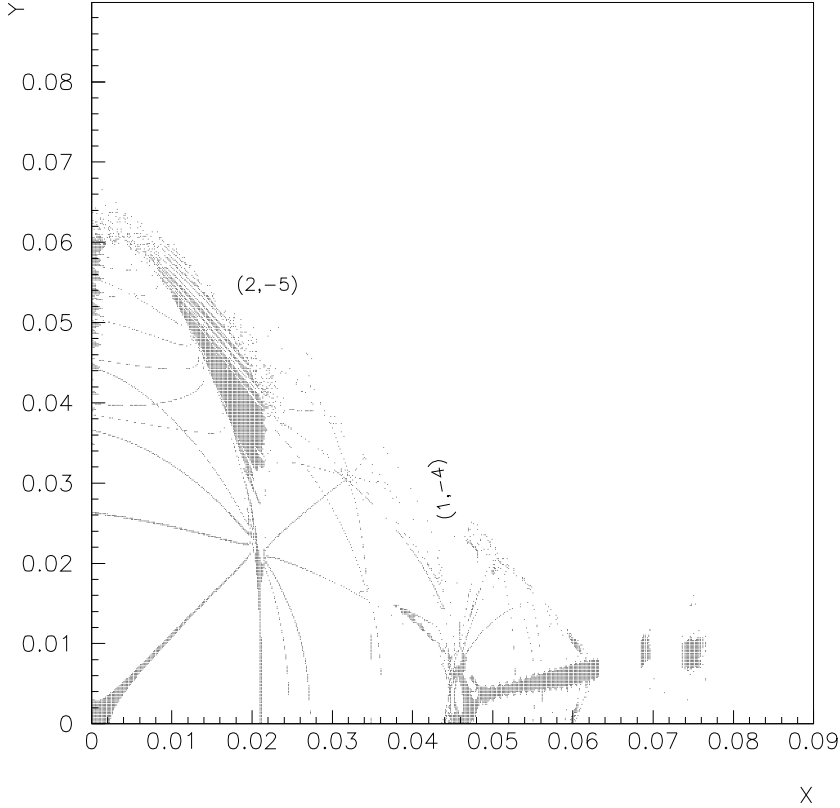, height=12truecm,width=12truecm}
\caption{Network of resonances after SR compensation.}
\end{figure}   

\subsection{Dynamic aperture calculations}
The dynamic aperture as a function of the number of turns $N$ can be defined
\cite{dyn1} as the first amplitude where particle loss occurs, averaged over the
phase space. Particle are started along a grid in the physical plane $(x,y)$:
\begin{equation}
x=r\cos\theta ~~~~~~~~~~~~~y=r\sin\theta
\end{equation}
and initial momenta $p_x$ and $p_y$ are set to zero.
\\
Let $r(\theta,N)$ be the last stable initial condition along $\theta$ before the
first loss (at a turn number lower than $N$). The dynamic aperture is defined as
\begin{equation}
D=\left[\int_0^{\frac{\pi}{2}}[r(\theta,N)]^4\sin(2\theta)\mbox{d}\theta\right]
^{\frac{1}{4}}.
\end{equation}
\\
An approximated formula for the error associated to the discretization 
both over the radial and the angular coordinate 
can be obtained replacing the dynamic aperture definition with a simple average
over $\theta$. Using a Gaussian sum in quadrature the associated error reads
\begin{equation}
\label{errore}
\Delta D=\sqrt{\frac{(\Delta r)^2}{4}+\left\langle\left|\frac{\partial r}
{\partial\theta}\right|\right\rangle^2\frac{(\Delta \theta)^2}{4}}
\end{equation}
where $\Delta r$ and $\Delta \theta$ are the step sizes in $r$ and $\theta$
respectively.
\\
In Tab. 2 the values (with the associated errors) of the dynamic
aperture are quoted for the three studied optics for short ($N$=5000) and 
medium ($N$=20000)
term tracking.
\begin{table}[h!]
\begin{center}
\vspace{5mm}
\begin{tabular}{|c| |c| |c| |c| }
\hline 
$D$ (m)&Uncoupled&Summed&Single\\
\hline\hline
$N$=5000&0.0406&0.0412&0.0372\\
\hline
$N$=20000&0.0405&0.041&0.037\\
\hline
\end{tabular}
\end{center}
\caption{Dynamic aperture values relative to the uncoupled Henon map and after
the summed and single resonance compensations. The associated error (according to the
formula (\ref{errore})) is about 2\% for $N$=5000 and about 4\% for $N$=20000.}
\end{table}

The difference between the summed and the single resonance compensations is noticeable: the
compensation of the all families relative to the coupling resonances allows
an improvement close to 10\% respect to the case in which the high frequency
part of the perturbative hamiltonian is neglected.
\\
It can also be pointed out that the summed compensation seems to slightly improve
(at the limit of sensitivity due to errors) the dynamic aperture respect to the
uncoupled case.

\section{The nonlinear case}
\label{extnl}
Dealing with high-order resonances with a view to optimizating stability is
not so straightforward as in the linear case: the number of resonances that can
be excited both by a given multipole and by the set of correctors meant for
compensating a given resonance becomes higer and higher; moreover the 
resonance compensation
is only one of the tools that has to be used to get a succesfully optics 
optimization.
\\

For these reasons we have not here attempted a general comparison between 
the summed and the
single resonance compensations using tracking analysis. We intend to return 
to this question in the future.
\\
We note however that a certain number of attempts to compensate one particular sextupolar
resonance for the Henon map show that the two compensations are not far only if
the working point is close enough to the considered resonance. The summed
resonance compensation is to in general better in the case of the 
compensation of several resonances at
the same time.

\newpage
\section{Conclusions}
\label{concl}
A general method has been derived for the summation of all the resonances within
a given family both for the linear and for the non-linear cases. The fact that
this summation is valid and gives a meaningful result is confirmed by its
application to the known closed-orbit distortion equation, the betatron
modulation equation and the decoupling of the linear transfer matrix for a ring.
The application of the summed-resonance driving term to the coupling raises the
question of the relative merits of the different types of coupling compensation
that are now possible. This problem has been investigated with the help of the Henon
map. The results indicate that use of the summed-resonance compensation 
(equivalent to
the matrix approach) yields a larger dynamics aperture.

\begin{center}
\Large
{\it Acknowledgements}
\end{center}
The work of D. Fanelli is supported by a Swedish Natural Science Research Council 
graduate student fellowship. We thank P.J. Bryant and  E. Aurell for 
discussions and critical reading of the manuscript.
\\
\\
\\

\newpage

\vspace{3truecm}

$\mbox{\bf {Figures caption:}}$

\vspace{3truecm}

$~~~\bullet$~~Figure 5: Stability domain of the uncoupled Henon map.

\vspace{1truecm}

$~~~\bullet$~~Figure 6: Footprint diagram of the uncoupled Henon map.

\vspace{1truecm}

$~~~\bullet$~~Figure 7: Stability domain after the summed resonance
                        compensation.  
\vspace{1truecm}

$~~~\bullet$~~Figure 8: Footprint diagram after the summed resonance
                        compensation. 
\vspace{1truecm}

$~~~\bullet$~~Figure 9: Stability domain after the single resonance
                        compensation.  
\vspace{1truecm}

$~~~\bullet$~~Figure 10: Footprint diagram after the single resonance
                        compensation.

\end{document}